\documentclass[12pt]{iopart}
\usepackage{epsf,epsfig,pstricks,graphicx}
\begin{document}
\title{Conductance through atomic point contacts between 
fcc(100) electrodes of gold} 
\author{O. Lopez-Acevedo$^1$, D. Koudela$^1$ and H. H\"akkinen$^{1,2}$}
\address{$^1$ Nanoscience Center and Department of Physics, FI-40014 University of Jyv\"askyl\"a, Finland}
\address{$^2$ Department of Chemistry, FI-40014 University of Jyv\"askyl\"a, Finland}
\begin{abstract}Electrical conductance through various nanocontacts between
gold electrodes is studied by using the density functional theory, scalar-relativistic 
pseudopotentials, generalized gradient approximation for the exchange-correlation
energy and the recursion-transfer-matrix method along with channel 
decomposition. The nanocontact is modeled with pyramidal  fcc(100) tips and
1 to 5 gold atoms between the tips. Upon elongation of the contact by adding gold atoms 
between the tips, the conductance at Fermi energy $E_F$ 
evolves from $G\approx 3 G_0$ to $G\approx 1 G_0$ ($G_0 = 2e/h^2$). 
Formation  of a true one-atom point contact, with
$G\approx 1 G_0$ and only one open channel,  requires
at least one atom with coordination number 2 in the wire. Tips that share
a common vertex atom or tips with touching vertex atoms have three partially open conductance
channels at $E_F$, and the symmetries of the channels
are governed by the wave functions of the tips.
The long 5-atom contact develops conductance oscillations and conductance
gaps in the studied energy range $-3 \leq E-E_F \leq 5$ eV, which reflects oscillations
in the local density of electron states in the 5-atom linear "gold molecule"
between the electrodes, and a weak coupling of this "molecule" to the tips.
\end{abstract}
\ead{lopez@phys.jyu.fi}
\maketitle

\section{Introduction}

Stable atomic point contacts and chain-like
wires  between metal leads constitute 
an intensive research field aimed at atomic-precision manipulation
of mechanical and electrical properties of nanojunctions for possible
use in future nanoelectronics \cite{1,2}. 
Gold nanowires have been an archetypical system; early experimental
discoveries \cite{3,4}
of stable several-atom-long wires have sparked 
extensive theoretical and experimental interest \cite{5}-\cite{23}.
Along with detailed results for
the dependence of conductance on the nanowire structure, early electronic structure calculations predicted also considerable chemical "sensing activity" of these contacts \cite{9,10}.

Gold nanowires are ideal systems to study ballistic transport in an
atomic environment, since (i) the wires are shorter than the electron
mean-free path and (ii) the Fermi wave length of gold is comparable to
the diameter  of an ultimately narrow wire (one gold atom), allowing for passage
of current with one unit of the quantum conductance $G_0 = 2e^2/h$
at energies close to $E_F$.
Step-wise reduction of the conductance down to $G\approx 1 G_0$
upon stretching (and concomitant narrowing) of the contact  
is well documented in many experiments and calculations.
Specifically, formation of several-atom long wires with $G\approx 1 G_0$
was reported in the ground-breaking experiments in 1998
\cite{3,4}.
Particularly in the mechanical break-junction experiment
\cite{4} formation of a one-atom point contact was attributed
to the discrete jump of conductance down to $G\approx 1 G_0$. 
This result has been conventionally interpreted as there being
just one open conductance channel (s-like channel,
according to the monovalent nature of gold \cite{5}).

This work re-iterates the fundamental question: when does a wire
constitute a "truly"  one-atom point contact with only single  open channel?
On the basis of density-functional-theory (DFT) calculations of relaxed
model structures and accompanied recursion transfer-matrix (RTM) calculations
for the (intrinsic, without bias) conductance, we show that a conductance
value close to the quantum unit is not a sufficient criterion to identify
a "true" atomic point contact between the leads, modeled here with
two fcc(100) tips. Upon formation of the wire, 
configurations exist which appear to have a minimal cross-section of just one gold atom,
however up to three conductance channels can be partially open to yield
a total conductance $G\approx 1 G_0$.  Formation of a true one-atom point
contact with only one open conductance eigenchannel
 requires at least one atom in the wire with a coordination number of two,
i.e., {\it three atoms} in a linear configuration. This way, the
formation of the point contact is intimately linked to the chemical
coordination of the atoms {\it in the wire}.
Tips that share
a common vertex atom or tips with touching vertex atoms have three partially open conductance
channels at $E_F$, and the symmetries of the channels
are governed by the wave functions of the tips.
We have also studied a long 5-atom wire (7 atoms altogether in a linear
configuration) and observed that it
 develops conductance oscillations and conductance
gaps in the studied energy range $-3 \leq E-E_F \leq 5$ eV. This reflects oscillations
in the local density of electron states in the 5-atom linear "gold molecule"
between the electrodes.

\section{Method}
Atomic structures of the model contacts 
 were relaxed and the electronic structure
analysed by using a density-functional total-energy method based
on ref. \cite{24}.  
The interactions between the 5d
 and 6s electrons and the core electrons of gold were described
by scalar-relativistic
Troullier-Martins norm-conserving first-principles pseudopotentials
\cite{25} in conjunction with the Perdew-Burke-Ernzerhof
(PBE) exchange-correlation potential \cite{26}.
 The nanocontact is built of two fcc(100) tips with
a variable number of gold atoms between the tips in the (100)-axis (z), see
Fig. 1. The length of each contact was relaxed with the constrain that the
atoms in the  outermost 
layers of the tips were fixed to their lattice-terminated (x,y) positions.

For obtaining the conductance, the local part of the self-consistent
Kohn-Sham potential $U_{KS}^{loc}({\bf r})$ of the DFT calculation was used in
the recursion-transfer-matrix
method along with the conductance eigenchannel
decomposition \cite{27,28,29}.
The transmission of the electron propagating from one electrode to another
through $U_{KS}^{loc}(\bf r)$ of the wire region is evaluated using a numerical
solution of the stationary states of the Schr\"odinger equation. In the direction
of the wire, propagation of the electrons is started from a smooth jellium potential
(at -11 eV) of
the lead and stopped when the bulk value of the potential is reached on the
opposite lead. In the transverse direction periodic boundary conditions are used.

The conductance $G(E)$ was calculated from the Landauer-B\"uttiker formula
as a sum over all non-zero conductance channels
\begin{equation}
 G = G_0 \Sigma_i \vert \tau_i\vert^2
\end{equation}
where $\vert \tau_i\vert^2$ are the eigenvalues of the product of transmission
matrices, $t^\dagger t$.
Convergence of the obtained conductance values $G(E)$
has been carefully tested with respect to number of plane
waves in the transverse direction, number of atomic layers in the tips,
number of points in the calculation grid into the bulk region of the lead. Finally, we have checked that the use of the local 6s-part for constructing the  $U_{KS}^{loc}$ from the non-local self-consistent Kohn-Sham potential that includes the 5d-6s-6p angular momentum channels  of the pseudopotential is a quantitatively satisfying approximation.

\begin{figure}
\includegraphics[scale=0.45]{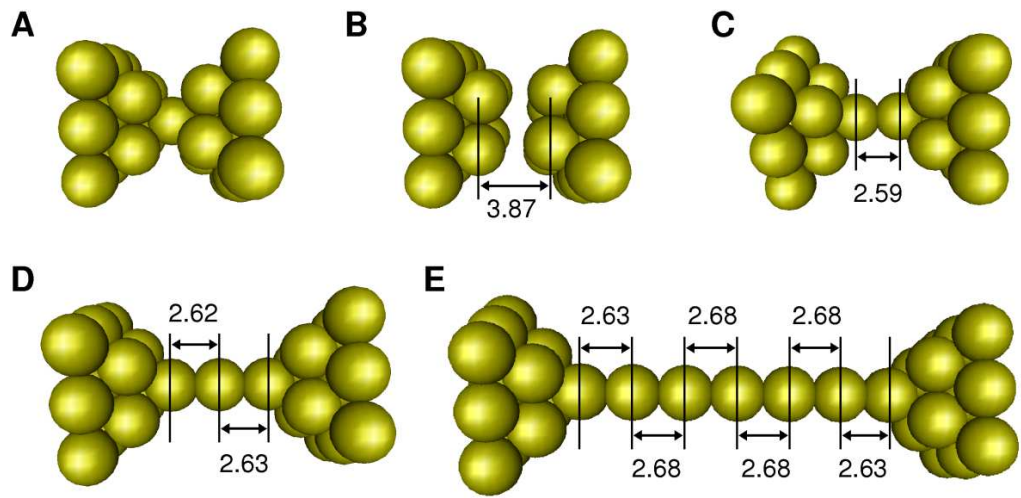}
\caption{Models for gold nanocontacts between fcc(100) tips:
A: tips sharing the vertex atom, B: as A, but the central atom removed
(non-relaxed), C: touching vertex atoms, D: one-atom contact, E:
 linear five-atom contact. Selected Au-Au distances are marked (in \AA).} 
\label{coords}
\end{figure}

\section{Results and discussion}
\begin{figure}
\includegraphics[scale=0.5]{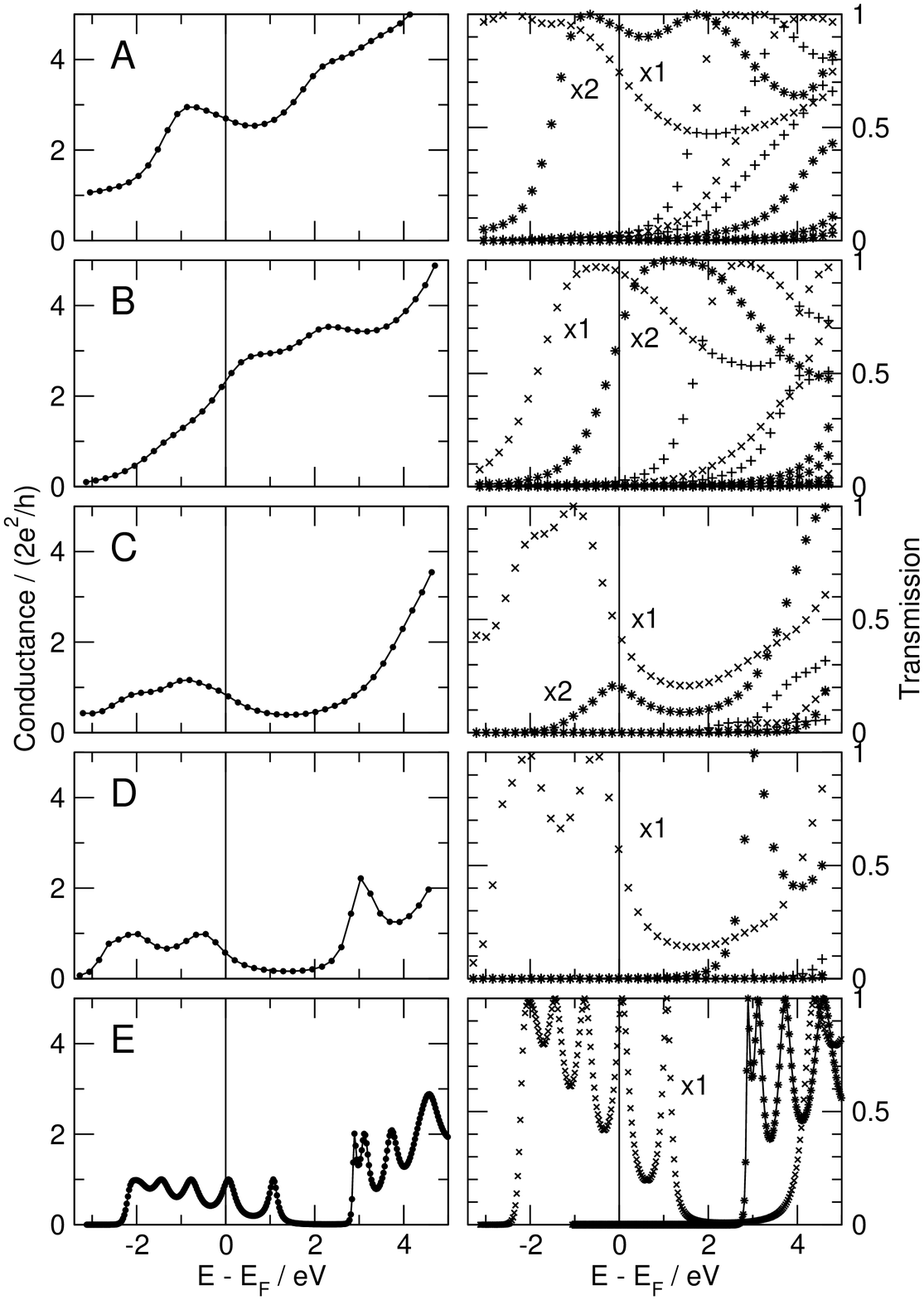}
\caption{Left: Conductance vs. energy for structures A - E of Fig. 1. 
	Right:  Transmission of eigenchannels. 
	The solid line marks the position of the Fermi energy. Degeneracies of the
	major channels indicated.}
\label{cond}
\end{figure}

Figure \ref{cond} shows conductance versus energy (left column)  and
the eigenchannel decomposition (right column) for all the configurations A-E shown
in Fig. 1. Table 1 summarizes the conductance values at the Fermi energy, $G(E_F)$.

\begin{table}
\caption{\label{tab}Conductance $G(E_F)$ and the number 
of open eigenchannels $N_C$ at the Fermi energy for contacts A - E. An
eigenchannel is considered as open if its transmission is larger than
0.05.}
\begin{indented}
\item[]\begin{tabular}{@{}lll}
\br        
        Contact & $G(E_F) / G_0$ & $N_C$\\
\mr        
        A & 2.7 & 3 \\
        B & 2.3 & 3 \\
        C & 0.8 & 3 \\
        D & 0.6 & 1 \\
	E & 0.9 & 1 \\
\br        
\end{tabular}
\end{indented}
\end{table}


The contact A 
shows a local maximum in the conductance ($G=3$ G$_0$) 
1 eV  below $E_F$. At the Fermi level the conductance is $G=2.7$ G$_0$ and 
reaches a local minimum at slightly higher energies. This high
value of conductance was also obtained in \cite{18} for the same system.
The eigenchannel analysis shows 
three partially open conducting channels. At $E_F$, the transmittance of the
first (non-degenerate) and second channel (doubly degenerate) are 0.75  and 0.93, 
respectively.  Higher channels show transmittance below 0.05.

This contact provides the first example where on structural grounds one could
argue that a one-atom point contact has been formed. From the point of view of
the electronic structure, this is not the case. Existence of three major channels
implies dominance of the tip electronic structure. As a straightforward test,
we removed the central atom in structure A (see structure B in Fig. 1)
and re-calculated the electronic structure,
$U_{KS}^{loc}$ and the conductance. Three major channels remain open at $E_F$
and the conductance remains high, $G(E_F) = 2.3 G_0$. 

\begin{figure}
\includegraphics[scale=0.6]{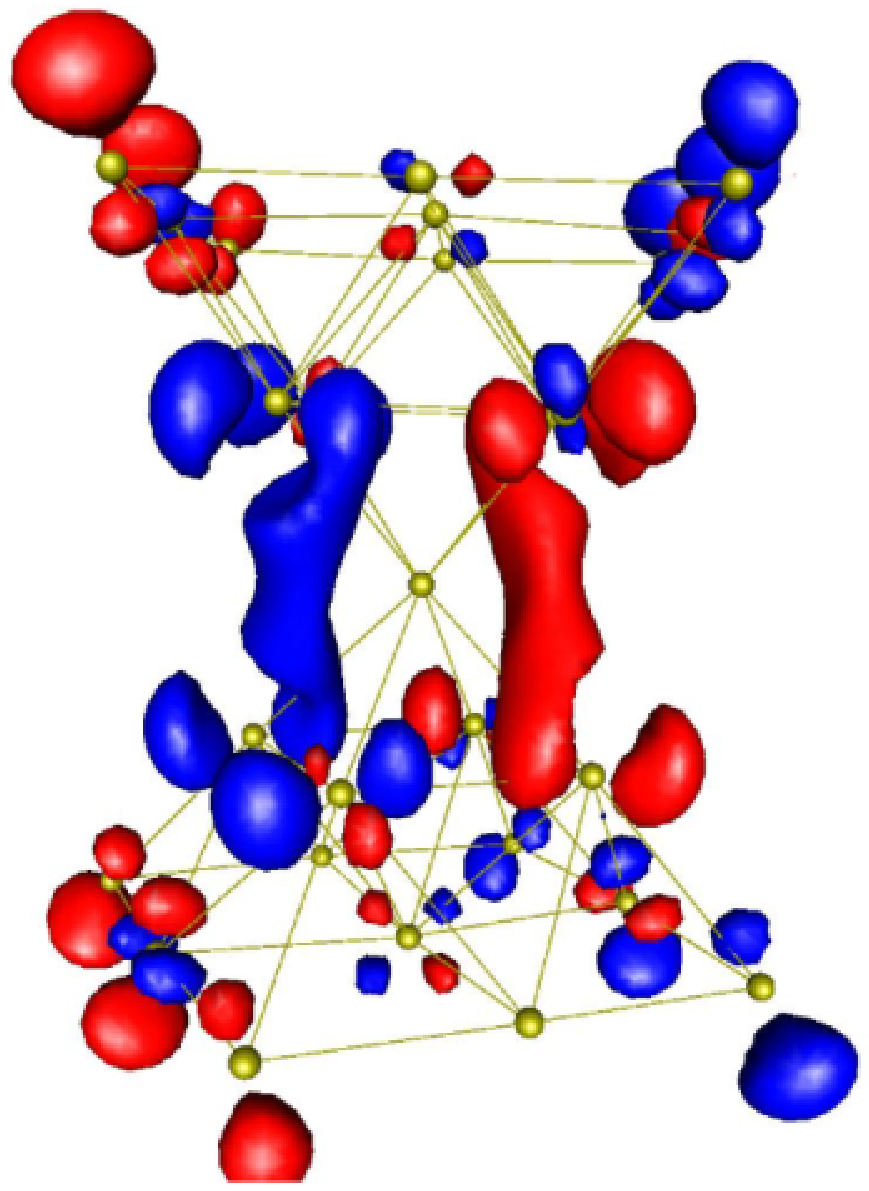}
\includegraphics[scale=0.6]{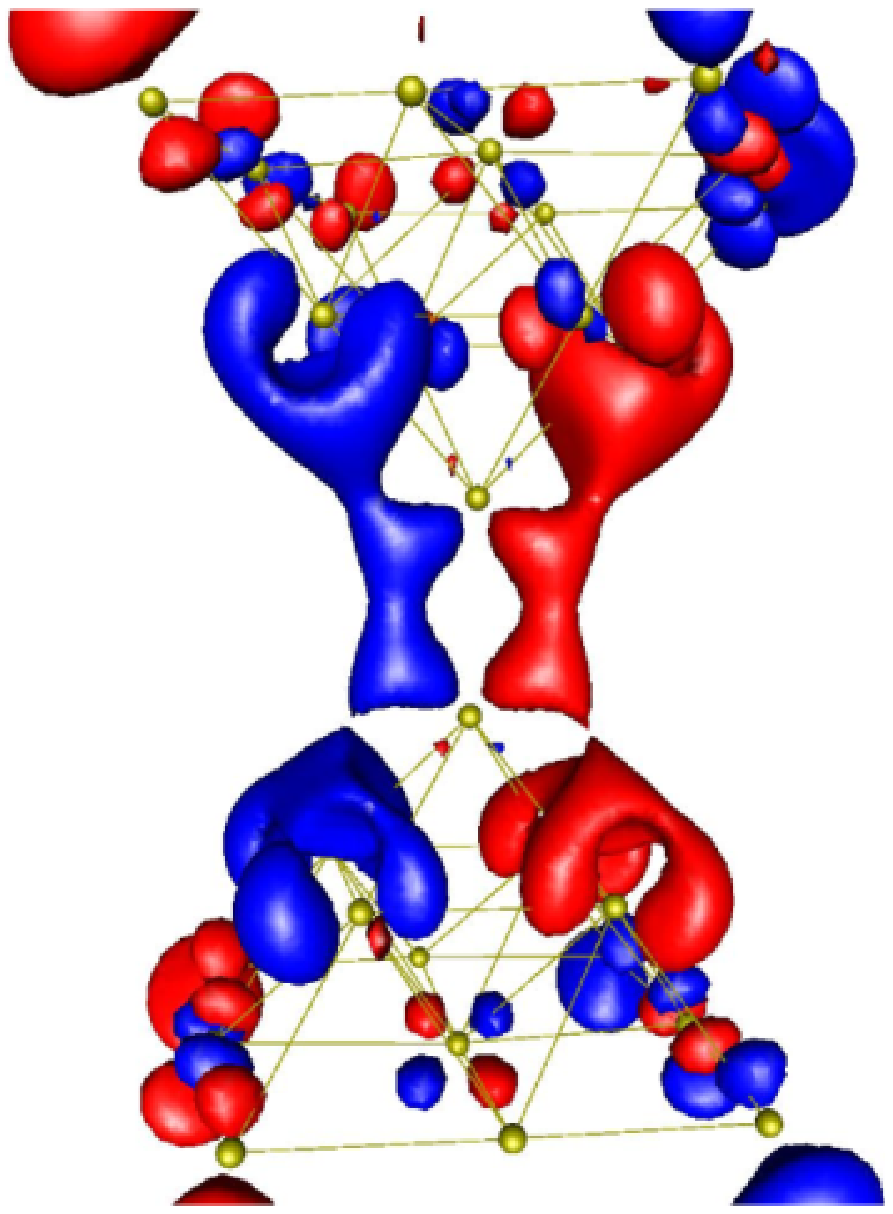}
\caption{ Left: The doubly degenerate Kohn-Sham orbital at $E_F$ for structure A. Note that
there is no weight around the shared vertex atom. Right: The corresponding
orbital for contact C.}
\label{wf1}
\end{figure}

As has been noted previously \cite{5}, the degeneracy patterns of conductance
channels close to Fermi energy can be related to symmetries of electronic orbitals in
the contact region. Fig. 3 (left) displays the visualization of one of the doubly-degenerate
 occupied orbital close to $E_F$ (the other one is related by a 45-degree rotation).
 It is interesting to note that this orbital has virtually no weight around the
 shared vertex atom. This  explains qualitatively also the insensitivity of the
 conductance to the removal of this atom (contact B).

$G(E_F)$ drops significantly (to 0.8 $G_0$) for contact C which consists of the tips with 
touching vertex atoms. The coordination number of the vertex atoms is 5. Based on the 
calculated conductance, one could again classify this contact as a "one-atom point contact".
However, the eigenchannel analysis reveals three partially open channels close to $E_F$.
The non-degenerate channel has a transmittance of 0.4 and the doubly-degenerate channel
0.2. Fig. 3 (right) displays a visualization of the doubly-degenerate orbital close to $E_F$,
which forms a bonding  state between  the touching vertex atoms, continuing all the way
to the tips.

Contact D has two maxima ($G\approx 1 G_0$) at  2 eV and
0.5 eV below $E_F$, and a rather low value at the Fermi level,
$G(E_F) = 0.6 G_0$. Only one channel remains open. It is interesting to correlate
the closing of the higher channels with the decrease of the coordination number
of the central atom from 5 (contact C) to 2 (contact D).  Also, the single open
channel at $E_F$ nearly closes for the region 1 - 2 eV above $E_F$.

The longest studied contact E (5-atom chain)  exhibits a qualitatively
distinct conductance pattern: oscillations  with sharp transmittance peaks
for a single eigenchannel at energies $-2 \leq  E-E_F  \leq 1 $eV and a wide conductance gap  
for
$1  \leq E-E_F \leq 3 $eV.   
The sharp peaks at energies  $3  \leq E-E_F \leq 5 $ eV  are  due to the
second eigenchannel which is doubly degenerate.

\begin{figure}
\includegraphics[scale=0.35,]{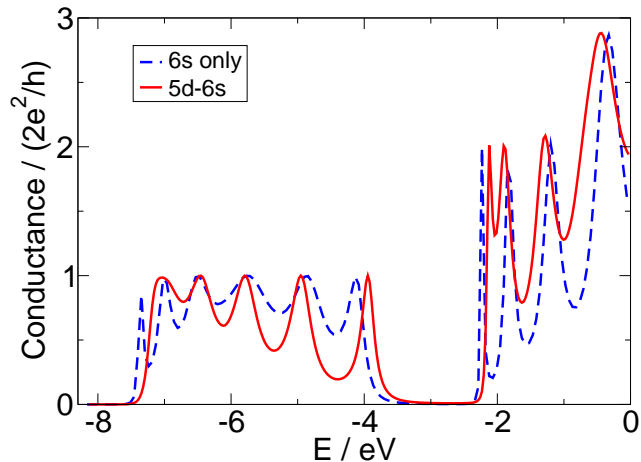}
\caption{Comparison between $G(E)$ calculated with the local part of the $5d-6s$ self-consistent pseudo-potential (red solid line) and $G(E)$ calculated with the local $6s$ self-consistent pseudo-potential (blue dashed line).} 
\label{cond-5}
\end{figure}

\begin{figure}
\includegraphics[scale=0.5,]{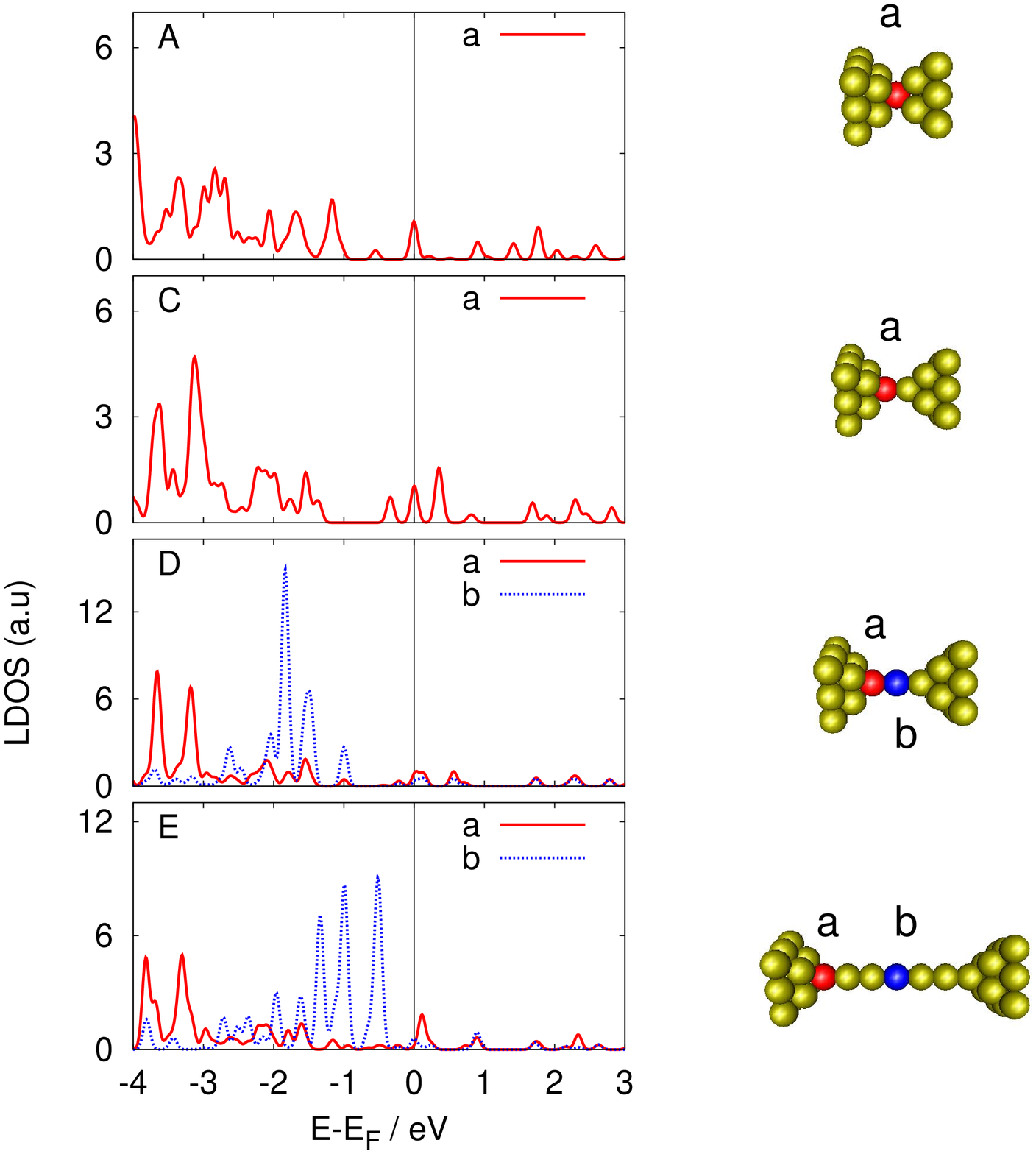}
\caption{ Left: Local density of electronic states (LDOS) for contacts
A, C, D and E. Right: the corresponding atomic configurations. The colours of
the LDOS curves for the "a" and "b" atoms correspond respectively to the colouring of the atoms on the right. Note the different y-scales} 
\label{ldos}
\end{figure}

We have checked that the distinct behaviour of the contact E is not due to
approximations made in building the effective local potential $U_{KS}^{loc}$
for the RTM calculations. A direct evidence is shown in Fig. 4  where
comparison of two $G(E)$ curves for contact E are shown.  The red
solid curve reproduces the data of Fig.  2  and the blue dashed curve
 is obtained from a RTM calculation where the $U_{KS}^{loc}$
 is derived from 6s-only pseudopotential DFT calculation for the contact.
 The shapes of the two conductance curves are very similar.
 
 What drives the observed distinct behaviour of contact E?
 One may note that a qualitative change of $G(E)$ takes place
 already when comparing contacts A and C: the direction of the wire axis
 $\lbrack 110\rbrack$ differs from the orientation of the (100) tips.
 This change is correlated with the development of a broad
 conductance minimum at energies $ 1\leq E-E_F \leq 2$ eV which develops
 further to a true conductance gap for contact E. 
 The $G(E)$ behaviour of contact E, a series of sharp "resonance" peaks
 at $-2 \leq  E-E_F  \leq 1 $ eV 
 separated by large conductance gaps from below and above, is a signature
 of a "molecular" nature of the contact. This is due to two factors:
 (i) the local electronic structure of the longer Au chains differs markedly
 from that in the tips, see Figure 5, which shows well-developed oscillations
 in the local density of electronic states projected to the central Au atom in the wire, and (ii)  mismatch of the crystal orientations between the wire and the tips ($[110]$ and $[100]$, respectively) causes significant back-scattering at the vertices of the tips, thus effectively decoupling the "molecular" Au chain from the tips. 

The mechanical stability of the long wire has been checked by modifying the linear five atom contact and optimizing. The modification consisted in a random shift of the wire atom coordinates in the plane perpendicular to the initial linear chain. After optimization a new isomer with lower energy has been found with the five central atoms in a zig-zag final configuration.  

\section{Conclusions}
We have studied the conductance of different one-atom-size contacts
between fcc(100) gold tips. Formation of a true one-atom point contact,
with $G\leq 1 G_0$ and one open eigenchannel, requires at least one atom
in the wire with a coordination number of two. The long 5-atom contact develops
sharp resonances and conductance gaps, signaling formation of a weakly
coupled "linear gold molecule" between the tips. The weak coupling
may be due to the fact that the linear "molecular" part of the contact
has locally a $[110]$ axis whereas the tips are in the $[100]$
orientation, causing significant back-scattering at the vertex atoms.
 Experimental observation
of this phenomenon might however be difficult due to various possible
non-ideal atomic configurations of the tips. At the peaks of the resonances, 
the long linear chains appear to be ideal one-channel ballistic conductors
with $G\approx 1 G_0$
close to $E_F$.   The disappearance of the experimentally measured
conductance for long contacts \cite{4}  is then truly due to mechanical instabilities
of the wire causing breaking of the contact, and not due to 
inherent deterioration of conductance in long linear wires.

\ack
We thank Matti Manninen and Michael Walter for fruitful discussions. 
This work is supported by the Academy of Finland via project 110013.
Computations were performed in the Nanoscience Center of the University of Jyv\"askyl\"a and in the Finnish IT Center for Science (CSC) in Espoo. 



\end{document}